# Apollonian emulsions

Sylvie KWOK[1*], Robert BOTET[2], Abdulwahed SHGLABOW[1], Bernard CABANE[1]

22nd May 2020

We have discovered the existence of extremely polydisperse High Internal-Phase-Ratio Emulsions (HIPE) in which the internal-phase droplets, present at 95% volume fraction, remain spherical and organize themselves in the available space according to Apollonian packing rules. Such Apollonian emulsions are obtained from dispersing oil dropwise in water in the presence of very little surfactant, and allowing them to evolve at rest for a week. The packing structure of the droplets was confirmed through size distribution measurements that evolved spontaneously towards power laws with the known Apollonian exponents, as well as comparison of the structure factors of aged HIPEs measured by Small-Angle X-ray Scattering with that of a numerically simulated Random Apollonian Packing. Thanks to the perfect sphericity of the droplets, Apollonian emulsions were found to display Newtonian ow even at such extremely high volume fraction. We argue that these fascinating space-filling assemblies of spherical droplets are a result of coalescence and fragmentation processes obeying simple geometrical rules of conserving total volume and sphericity, minimizing the elastic energy associated with interactions of neighbouring droplets.

## 1. Historical introduction

In the third century B.C., in Hellenistic Alexandria, lived Apollonius of Perga who wrote ten famous books on geometry[1]. In his book "*Επαφαι*" ("Tangencies"), the last and most challenging problem was the construction of circles that would be tangent to three given circles in the Euclidean plane. All copies of this book are now lost, but the question survived as an intellectual challenge, thanks to Pappus of Alexandria who reviewed the problem 500 years later and named it "Apollonius' circle problem"[2]. From there, it became enriched through centuries of discussions amongst mathematicians: Viète (1600) reconstructed the plausible Apollonian solution using clues from ancient literature[3]; Descartes (1643) introduced the premise of three mutually tangent circles at the initial stage[4]; Leibniz (1706)[5] envisioned packing the plane with an infinity of non-overlapping circles; Kasner et al. (1943)[6,7] proved Leibniz's conjecture, and Mandelbrot (1977) established the fractal nature of such a packing[8]. In parallel, Apollonius' circle problem was extended to the three-dimensional space, with Soddy (1937) introducing close-packing in the context of an "infinitely infinite number of spheres that theoretically can be packed into [a hollow sphere]"[9]. About real applications of this problem to our world, Newton (1687) recognized its equivalence with triangulating a position from the differences of its distance with respect to three known points, giving us the ubiquitous Global Positioning System[10]. Medical imaging[11,12] and pharmacology[13] have similarly benefited from Apollonius, both in experiment and simulation work. This non-exhaustive list demonstrates the problem's timelessness, and how it has occupied some of the most brilliant minds through the ages.

## 2. Outline of the present contribution

We consider here the three-dimensional variant of the Apollonian packing problem: by repeating the procedure ad infinitum, one could generate an infinite number of increasingly smaller non-overlapping spheres that would completely fill the spaces between the original ones. The final attainable volume fraction is thus 100%, contrary to the case of monodisperse spheres where it is 74%. Since each constructed sphere must be selected precisely for its diameter, a constraint is set on the diameters of particles that will be used to fill the voids. If the original spheres are non-identical, the constructed spheres must possess a continuous power-law distribution of diameters[14]. The exponent of this power law is $-(d_f + 1)$, where $d_f$ is the fractal dimension of the union of all sphere surfaces in the dense packing[8]. Advancements in numerical simulations have enabled accurate determination of the Apollonian packing exponent, usually found to be $d_f \simeq 2.47$.[15–18]

Consequently, this recursive packing protocol should be useful for assembling spherical particles into materials with infinitely small porosity. However, the practical use of this protocol in material manufacturing is severely limited, due to the necessity of mixing discrete populations of spherical particles and hoping that each particle would manage to find its preferred place in the network. Given that small grains (colloidal particles or emulsion droplets) have a tendency to segregate during processing[19], it seems highly unlikely that the desired self-positioning can occur on its own. It would obviously be much more attractive and convenient if we could generate materials which spontaneously evolved the necessary size distributions and local placements. Here, we report a first step towards this goal, namely: the fabrication of emulsions that evolve spontaneously to become materials with a random Apollonian packing structure.

[1] Laboratoire Colloïdes et Matériaux Divisés - Chemistry, Biology and Innovation (CBI) UMR8231, ESPCI Paris, CNRS, PSL Research University, 10 rue Vauquelin, Paris, France
[2] Laboratoire Physique des Solides, CNRS UMR8502, Université Paris-Saclay, 91405 Orsay, France
* sylvie.kwok@espci.fr

## 3. The recipe for an Apollonian emulsion

We make these Apollonian emulsions by dispersing oil dropwise under constant stirring in an aqueous surfactant solution, much like making a mayonnaise, then allowing them to evolve at rest for at least a week. When the internal volume fraction ϕ exceeds the close-packing limit of spheres (64% for random[20] or 74% for ordered[21]), the emulsion is called a "High Internal-Phase-ratio Emulsion" (HIPE)[22]. In order to obtain space-filling structures, we continue the addition of oil until its volume fraction as the internal phase is extremely high (ϕ = 95% – 99%).

The behaviours of a HIPE, like those of any emulsion, are determined by the properties of the surfactant monolayers that are available to keep oil droplets separated. In a surfactant-rich HIPE (typically 7 – 30%wt surfactant in the continuous phase for non-ionic surfactants), strong monolayers (or multilayers) keep oil droplets separated as soon as emulsification is completed. Consequently, high ϕ can only be accommodated by deforming the oil droplets into polyhedral volumes separated by flat surfactant films [23–25] (Fig.1, left). The system then behaves like an elastic solid that withstands external stress by storing energy through further film distortion[26,27]. Such surfactant-rich HIPE have been extensively studied, and the shapes of the polyhedral droplets have been determined. However, we took a seemingly unexplored path by severely reducing surfactant availability to only 0.6% in the continuous phase (water). The critical micelle concentration of the non-ionic surfactant used here is 87µM, i.e. 0.004%wt in water. (The detailed protocol is described elsewhere[28]). We found that in surfactant-poor HIPEs, high ϕ is accommodated by an Apollonian packing of extremely polydisperse droplets: indeed, spaces between large droplets are filled at all scales by smaller ones (Fig.1, right), indicating that the droplets are packed to minimize elastic energy preferentially.

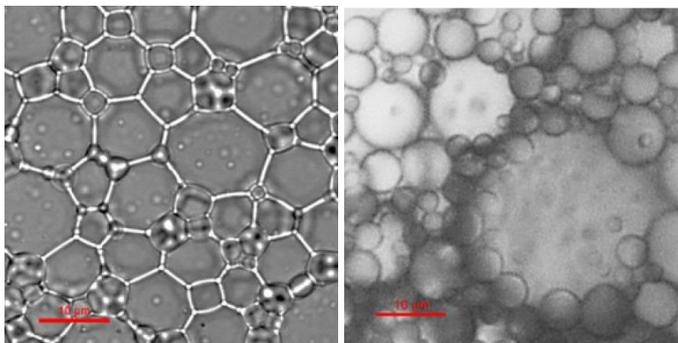

Figure 1: Optical microscope photos of oil-in-water HIPEs at ϕ = 95%. Depth of field around 5µm. (left) Surfactant-rich HIPE containing 20% non-ionic surfactant in the continuous phase. The droplets are distorted into polyhedral shapes. (right) Surfactant-poor HIPE made with only 0.6% surfactant. The droplets remain perfectly spherical, with an extremely broad size distribution.

## 4. Droplet-size distribution in Apollonian emulsion

Besides visual observation, another indication that our surfactant-poor HIPEs are Apollonian comes from measuring their droplet-size distributions through Light Scattering. At short times, their number-weighted diameter distributions could be characterized by power-law decays, with exponents ranging between –3 and –3.5. With time (about a week), these exponents consistently converged towards and remained between –3.48 and –3.50 regardless of the initial conditions (Fig.2). This range of values corresponds to $d_f$ = 2.48–2.50, in good agreement with reported Apollonian $d_f$.

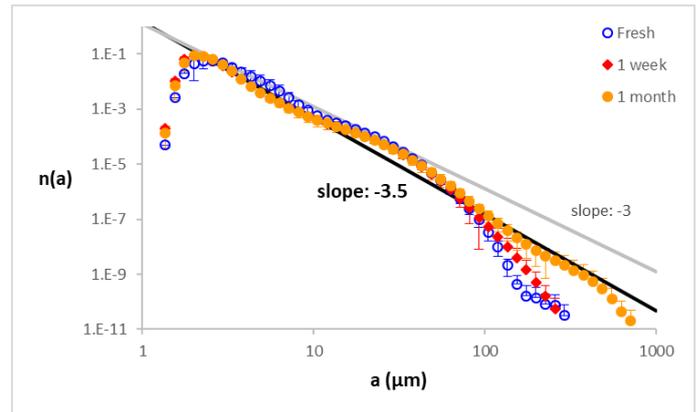

Figure 2: Evolution of droplet-diameter distribution of surfactant-poor HIPEs at ϕ = 95% measured by Light Scattering. The points shown are averaged over 3 samples, made at 200rpm, 500rpm and 1000rpm. Immediately after emulsification ended (open circles), the samples' size distributions followed a power law of exponent between -3 and -3.5. After 1 week of evolution at rest (solid diamonds), they then gradually evolved towards and stabilized at the Apollonian exponent (represented by the solid black line). Even after 1 month of evolution (solid circles), the power-law exponent remained unchanged.

## 5. Rheology of an Apollonian emulsion

Contrary to surfactant-rich HIPEs which exhibit plastic shear-thinning deformation beyond their yield stress[25,27,29], surfactant-poor HIPEs flow like Newtonian fluids (viscosity independent of shear rate, see Fig.3). Newtonian behaviour has never been documented in emulsions beyond ϕ = 60% [30–32], making it all the more surprising that a HIPE at 95% displayed such behaviour.

The absence of a yield stress in an Apollonian emulsion indicates that the system is below jamming, and the Newtonian flow implies that individual oil droplets in the Apollonian emulsion behave as frictionless particles would in an ideal hard-sphere suspension[33]. We may therefore infer that each droplet moves as if its neighbours were absent in spite of the extremely high volume fraction in an Apollonian emulsion. We conjecture that this is in part due to the perfect sphericity of the droplets where they may roll freely over one another without torsion friction, such as in space-filling bearings[34]. More importantly, studies of bi- and multi-modal suspensions have shown that if the size difference between sphere populations exceeds 10%, larger ones undergo a zig-zag motion; otherwise, they moved as if the ensemble of smaller spheres made up a homogeneous viscous fluid[30,35,36]. It would then appear that in a scale-invariant Apollonian emulsion where droplet-size distribution is a continuous power law, friction between droplets is quite low[37], and each smaller size-class of droplets plays the role of a homogeneous viscous medium to their larger neighbours, thereby enabling Newtonian flow.

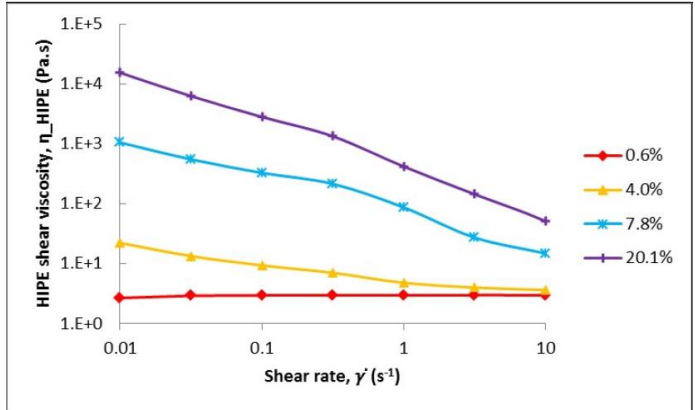

*Figure 3: Flow curves of HIPEs with different amounts of surfactant in the continuous phase. The Apollonian HIPE containing 0.6% non-ionic surfactant in the continuous phase displays Newtonian flow behaviour, as evidenced by the flat slope of the flow curve (red diamonds). As surfactant concentration increases, shear-thinning flow behaviour is observed, which manifests as a negative slope. This is the due to the progressive formation of facetted surfaces between distorted droplets.*

## 6. Structure factor of Apollonian emulsion

Finally, we measured the average correlation of oil droplet positions, using Small Angle X-ray Scattering (SAXS). This average correlation was characterized by an experimental structure factor, $S(q)$, which is the ratio of the intensity $I(q)$ scattered by the HIPE, to the intensity $P(q)$ scattered by the same droplets in a diluted HIPE (dilution ratio 1/40).[38] The experimental structure factors, $\langle S(q) \rangle$, of our surfactant-poor HIPEs are presented in Fig.4. In our HIPEs, the main peak had collapsed to $S_{max}$ = 1.1 with little oscillations, implying almost complete positional disorder among neighbouring droplets.

This peak collapse and lack of oscillations are significant to the interpretation of the spatial correlation of oil droplets: typically, $S(q)$ of moderately polydispersed systems found in literature present $S_{max}$ ≈ 2- 3 with dampening oscillations at higher q values. These are signs of long-ranged order; indeed, the stronger and the shorter-ranged the spatial correlations, the higher the $S_{max}$ and oscillations. While it is possible to encounter S(q) where $S_{max}$ is close to 1 for systems that are very polydisperse,[24,38,39] reported systems possessed either lognormal[39] or Schulz[38] size distributions – very different from the power-law distributions of our Apollonian HIPEs with a defined exponent. These reported systems were also prepared at volume fractions not exceeding 53%[24], far lower than ϕ = 95% in our HIPEs.

In the domain of concentrated dispersions, Scheffold and Mason[40] showed that Vrij's solution to the Percus-Yevick approximation was valid and could accurately produce the spatial correlations of a densely packed disordered polydisperse emulsion up to ϕ = 72%. We therefore investigated if Vrij's solution could reproduce our experimental S(q) when our experimental parameters (size distributions and volume fraction) were fed into the calculations (Fig.4). For the sake of completeness, we also compared our experimental $\langle S(q) \rangle$ with that of numerically simulated spheres packed at similar volume fraction according to a 3D Osculatory Random Apollonian Packing (ORAP) algorithm[28]. A 3D ORAP is built by sequentially adding the largest sphere compatible with the remaining voids, under the condition of gentle tangency with neighbouring spheres delimiting the void. An example of an ORAP system is shown in Fig.5.

We found that Vrij's solution, which only assumes disorder and does not assume a particular droplet packing, overestimated our experimental $S_{max}$. On the contrary, the ORAP-generated S(q) presented excellent agreement with our experiments, indicating convincingly that the droplets in surfactant-poor HIPEs are indeed organized like a Random Apollonian Packing.

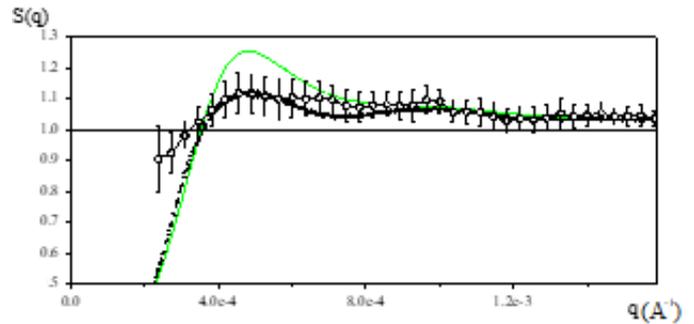

*Figure 4: Averaged experimental structure factor, $\langle S(q) \rangle$, over three independent surfactant-poor HIPEs, ϕ ≈ 95%, measured by SAXS (data are shown as circles; the error bars are the standard deviations for each value of q). The black dots are the calculated S(q) for a numerical simulation of a 3D Osculatory Random Apollonian Packing system at a similar volume fraction (part of the system is shown in Fig.5). The fine solid green line is the calculated S(q) from Vrij's solution to the Percus-Yevick approximation and it is valid for any disordered polydisperse concentrated system that does not assume any geometrical packing.*

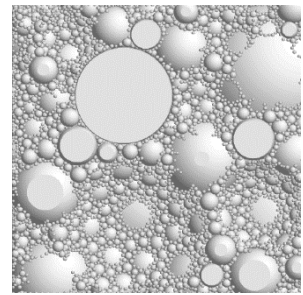

*Figure 5: Cross-section of a simulated ORAP system of 30,000 spheres, at ϕ = 92%. The S(q) shown as the solid curve of black dots in Fig.4 was obtained from this system.*

## 7. How is an Apollonian emulsion formed from a surfactant-poor HIPE?

Generally, the processes that occur during emulsion ageing at rest are the fusion of droplets (coalescence), their fission (fragmentation), and the exchange of oil molecules between neighbours in contact. In a common dilute emulsion, these processes take place spontaneously if they minimize the total free energy, of which the main contribution is the Helfrich curvature free energy integrated over the interface area of each droplet; coalescence and ripening contribute to the reduction of total interfacial area and lead eventually to macroscopic separation of oil and water. In a surfactant-poor HIPE, however, the main contribution to the total free energy is rather the elastic interaction energy in this overcrowded network. We have found two pieces of evidence that demonstrate this claim – firstly, all the droplets in a surfactant-poor HIPE are spherical in shape (Fig.1); secondly, we found an increase in the total interfacial area as the HIPE aged at rest, inferred directly from the asymptotic Porod's limit of the HIPE's measured SAXS intensity (Fig.6 bottom). This increase is indicative of the creation of small droplets as the HIPE evolves.

Accordingly, we argue that the processes causing two droplets to recombine in a surfactant-poor, highly crowded HIPE result in the formation of many spherical droplets, some larger, some smaller than the parent droplets. Repetitive evolution according to this **coalescence-fragmentation** mechanism then produces an Apollonian structure, as the smaller spherical droplets created occupy cavities in the vicinity of the coalescence event. By matching exactly the size of the cavity (condition of gentle tangency), the free energy of the HIPE is thereby minimized.

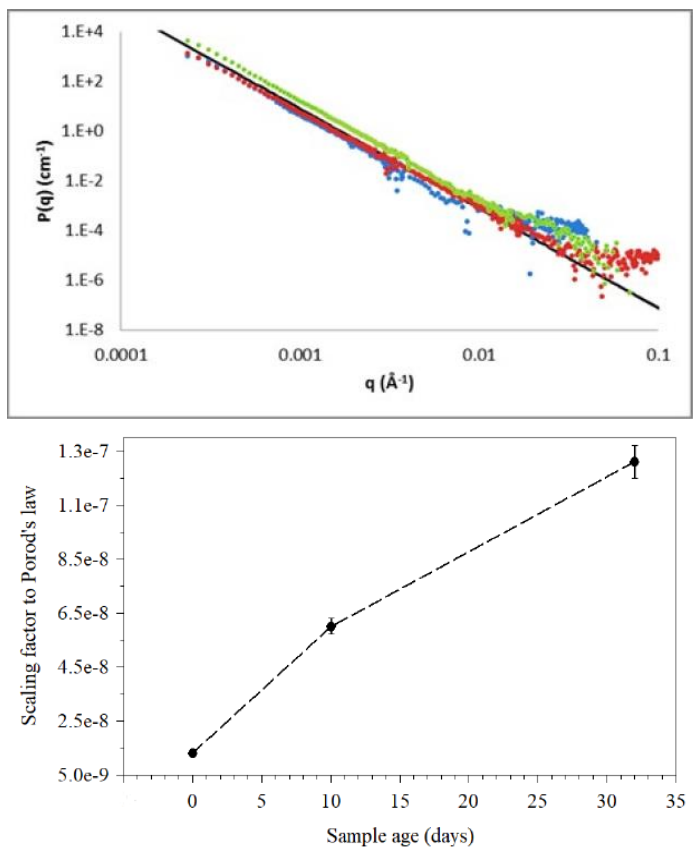

*Figure 6: (top) P(q) of surfactant-poor HIPEs measured by SAXS at different ages: freshly made (blue), 1-week-old (red), 1 month-old (green). The solid black line represents Porod's law; (bottom) the scaling factor, found by fitting the experimental data to Porod's law, is directly proportional to the total surface area of the matter scattering the X-rays. The major part of the error bars comes from background subtraction. The dashed line is a visual guide. This plot shows the regular increase in the total surface area of the same system over a month.*

We may then conclude that the main driving force of surfactant-poor HIPEs evolving towards an Apollonian structure is the minimization of elastic energy in these crowded structures by preserving sphericity. This coalescence-fragmentation mechanism differs from droplet coalescence in dilute emulsions where two droplets typically combine into a larger one. In the latter, the process is not spatially constrained and it allows for a change in droplet shape before concluding with a reduction in surface area and the subsequent expulsion of excess surfactant.

## 8. Possible extensions of the study

We made a cursory attempt to exploit the space-filling property of Apollonian emulsions by fabricating a solid material with little porosity and composed entirely with spheres. We used a polymer-solvent mixture of polycaprolactone and dichloromethane as the internal phase, and 0.02% ionic surfactant solution as the continuous phase. Applying our dropwise emulsification protocol, we attained $\phi$ = 92% and found an Apollonian packing of spherical liquid droplets (Fig.10, top). As the solvent evaporated, the polymer began to solidify[41]. We could therefore preserve the spherical geometry. However, non-uniform solvent evaporation from each Apollonian emulsion droplet resulted in somewhat imperfect sphericity in the final solid material (Fig.7, bottom). Nonetheless, this demonstrates the first steps towards the use of Apollonian emulsions for optimal fabrication of extremely porous or extremely dense solid materials. It is worth noting that solidification of the aqueous phase of an Apollonian emulsion would lead to an extremely porous solid structure of definite fractal dimension, $d_f \simeq 2.47$.

## 9. Conclusion

We have found a reproducible way to materialize an Apollonian packing of spheres by exploring surfactant-poor High Internal-Phase-Ratio Emulsions. It has remarkably taken over 2000 years since Apollonius of Perga defined the problem to discover a simple practicable condition that leads to such a fascinating structure in the real world. Our experimental observations show that droplets in such emulsions are spherical despite the very high internal-phase volume fraction, and their size distributions follow a power law consistent with the Apollonian fractal dimension. The droplets in these special HIPEs yield structure factors matching that of a numerically simulated Oscultatory Random Apollonian Packing. Such a kind of droplet organization has resulted in macroscopic behaviours differing from typical emulsions, for example, Newtonian flow, and the propensity to fill in the inner voids with tiny spherical droplets in spite of on-going coalescence.

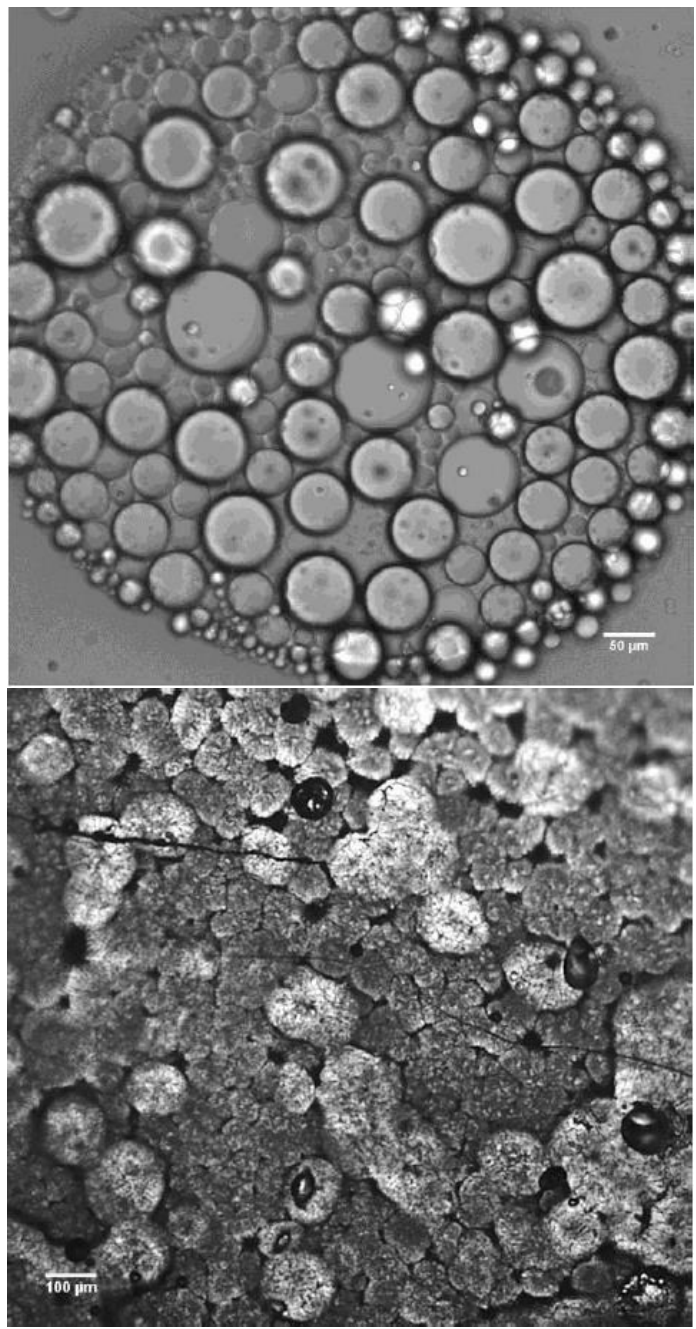

*Figure 7: (top) an Apollonian emulsion at ϕ = 92% made from a polymer-solvent mixture as the internal phase, and 0.02% ionic surfactant solution as the continuous phase; (bottom) as the solvent evaporated, the polymer solidified, preserving the space-filling Apollonian structure in the solid material obtained.*

**Acknowledgments**

The SAXS experiments were performed on beamline ID02 at the European Synchrotron Radiation Facility (ESRF), Grenoble, France and we are grateful to Lee Sharpnack for providing assistance. We also thank Yeshayahu Talmon, Lucy Liberman and Irina Davidovich of Technion Israel Institute of Technology for their help with microscopy. At last, we would like to thank Ryohei Seto (now at Wenzhou Institute, China) for mentioning to us the somewhat obscure Farris' effect; the work done by Farris ended up proving incredibly useful in interpreting many of our experimental observations.

*Mater. Sci. Eng.*, 2006, **14**, 789–798.